\documentstyle[epsf]{mn}

\newif\ifAMStwofonts

\ifoldfss
  \ifCUPmtlplainloaded \else
    \NewTextAlphabet{textbfit} {cmbxti10} {}
    \NewTextAlphabet{textbfss} {cmssbx10} {}
    \NewMathAlphabet{mathbfit} {cmbxti10} {} 
    \NewMathAlphabet{mathbfss} {cmssbx10} {} 
  \fi
  \ifAMStwofonts
    \ifCUPmtlplainloaded \else
      \NewSymbolFont{upmath} {eurm10}
      \NewSymbolFont{AMSa} {msam10}
      \NewMathSymbol{\upi}     {0}{upmath}{19}
      \NewMathSymbol{\umu}     {0}{upmath}{16}
      \NewMathSymbol{\upartial}{0}{upmath}{40}
      \NewMathSymbol{\leqslant}{3}{AMSa}{36}
      \NewMathSymbol{\geqslant}{3}{AMSa}{3E}

      \let\leq=\leqslant 
       
    \fi
  \fi
\fi 

\ifnfssone
  \newmathalphabet{\mathit}
  \addtoversion{normal}{\mathit}{cmr}{m}{it}
  \addtoversion{bold}{\mathit}{cmr}{bx}{it}
  \newmathalphabet{\mathbfit} 
  \addtoversion{normal}{\mathbfit}{cmr}{bx}{it}
  \addtoversion{bold}{\mathbfit}{cmr}{bx}{it}
  \newmathalphabet{\mathbfss} 
  \addtoversion{normal}{\mathbfss}{cmss}{bx}{n}
  \addtoversion{bold}{\mathbfss}{cmss}{bx}{n}
  \ifAMStwofonts
    \ifCUPmtlplainloaded \else
      \UseAMStwoboldmath
      \makeatletter
      \new@mathgroup\upmath@group
      \define@mathgroup\mv@normal\upmath@group{eur}{m}{n}
      \define@mathgroup\mv@bold\upmath@group{eur}{b}{n}
      \edef\UPM{\hexnumber\upmath@group}
      \new@mathgroup\amsa@group
      \define@mathgroup\mv@normal\amsa@group{msa}{m}{n}
      \define@mathgroup\mv@bold\amsa@group{msa}{m}{n}
      \edef\AMSa{\hexnumber\amsa@group}
      \makeatother
      \mathchardef\upi="0\UPM19
      \mathchardef\umu="0\UPM16
      \mathchardef\upartial="0\UPM40
      \mathchardef\leqslant="3\AMSa36
      \mathchardef\geqslant="3\AMSa3E

      \let\leq=\leqslant 

    \fi
  \fi
\fi 

\ifnfsstwo
  \DeclareMathAlphabet{\mathbfit}{OT1}{cmr}{bx}{it}
  \SetMathAlphabet\mathbfit{bold}{OT1}{cmr}{bx}{it}
  \DeclareMathAlphabet{\mathbfss}{OT1}{cmss}{bx}{n}
  \SetMathAlphabet\mathbfss{bold}{OT1}{cmss}{bx}{n}
  \ifAMStwofonts
    \ifCUPmtlplainloaded \else
      \DeclareSymbolFont{UPM}{U}{eur}{m}{n}
      \SetSymbolFont{UPM}{bold}{U}{eur}{b}{n}
      \DeclareSymbolFont{AMSa}{U}{msa}{m}{n}
      \DeclareMathSymbol{\upi}{0}{UPM}{"19}
      \DeclareMathSymbol{\umu}{0}{UPM}{"16}
      \DeclareMathSymbol{\upartial}{0}{UPM}{"40}
      \DeclareMathSymbol{\leqslant}{3}{AMSa}{"36}
      \DeclareMathSymbol{\geqslant}{3}{AMSa}{"3E}

      \let\leq=\leqslant 

    \fi
  \fi
\fi 

\ifCUPmtlplainloaded \else
  \ifAMStwofonts \else 
    \def\upi{\pi}
    \def\umu{\mu}
    \def\upartial{\partial}
  \fi
\fi

\def\eqb{\begin{eqnarray}}
\def\eqe{\end{eqnarray}}
\def\that{\hat{t}}
\def\nuf{\hat{\nu}}
\def\emf{\hat{\cal{E}}_{\hat{\nu}}}
\def\emo{{\cal{E}}_\nu}

\title[Relativistic blastwaves and synchrotron emission]{Relativistic 
blastwaves and synchrotron emission}
\author[T.\ P.\ Downes et al.]
       {T.\ P.\ Downes$^1$\thanks{Affiliated to the Dublin Institute for
Advanced Studies.}, P.\ Duffy$^2$ and S.S.\ Komissarov$^3$ \\
$^1$School of Mathematical Sciences, Dublin City University, Glasnevin,
Dublin 9, Ireland \\
$^2$Department of Mathematical Physics, National University of Ireland,
Dublin, Dublin 4, Ireland \\
$^3$Department of Applied Mathematics, University of Leeds, Leeds LS2
9JT, United Kingdom}
\date{Accepted ---.
      Received ---;
      in original form ---}

\pagerange{\pageref{firstpage}--\pageref{lastpage}}
\pubyear{2010}

\begin{document}

\maketitle

\label{firstpage}

\begin{abstract}

Relativistic shocks can accelerate particles, by the first order Fermi
mechanism, which then emit synchrotron emission in the post
shock gas. This process is of particular interest in the models used for the
afterglow of gamma ray bursts. In this paper we use recent results in the
theory of particle acceleration at highly relativistic shocks to model the
synchrotron emission in an evolving, inhomogeneous and highly
relativistic flow. We have developed a numerical code which integrates the
relativistic Euler equations for fluid dynamics with a general equation
of state, together with a simple transport equation for the accelerated 
particles.  We present tests of this code and, in addition, we use it to 
study the gamma ray burst afterglow predicted by the fireball model,
along with the hydrodynamics of a spherically symmetric relativistic
blastwave.

We find that, while, broadly speaking, the behaviour of the emission is 
similar to that already predicted with semi-analytic approaches, the detailed 
behaviour is somewhat different.  The ``breaks'' in the synchrotron
spectrum behave differently with time, and the spectrum above the final 
break is harder than had previously been expected. These effects are due 
to the incorporation of the geometry of the (spherical)
blastwave, along with relativistic beaming and adiabatic cooling of the
energetic particles leading to a mix, in the observed spectrum, between
recently injected "uncooled" particles and the older "cooled" population in
different parts of the evolving, inhomogeneous flow.

\end{abstract}

\begin{keywords}
gamma rays: bursts -- hydrodynamics -- radiation mechanisms: non-thermal
-- relativity
\end{keywords}

\section{Introduction}

Relativistic shock fronts arise in astrophysics when a
relativistic flow propagates into ambient material such as in the
jets of active galactic nuclei (AGNs) and fireball models of gamma
ray bursts (GRBs). These shock fronts are also sites of energetic
particle acceleration which in turn leads to non-thermal emission
processes such as synchrotron radiation and inverse Compton
scattering. Particle acceleration at relativistic shock fronts 
can occur via the first order Fermi mechanism. In
this picture, energetic particles of speed $v$ are scattered back
and forward across the shock front by magnetic turbulence. Each
crossing involves the particle meeting the flow {\it head-on}
leading to an increase of energy. At nonrelativistic shocks,
$v>>U_{1,2}$ where $U_1$ and $U_2$ are the upstream and downstream 
flow speeds in the shock rest frame respectively.  With diffusive 
particle scattering the particle distribution is almost isotropic both 
upstream and downstream.  This results in a power law index for the 
particle spectrum which only depends on the shock compression ratio, 
$r=U_1/U_2$. In relativistic flows the situation is more
complicated. When $v$ is comparable to $U_{1,2}$ the timescale for
isotropisation of the energetic particle population by the
magnetic turbulence (which determines the degree of anisotropy of the
particle distribution upstream and downstream) becomes important. 
Moreover even if the particle distribution were isotropic in one fluid frame 
it will be anisotropic when viewed in the shock or other fluid frame.
As a result, unlike the case for nonrelativistic flows, the
particle spectrum depends not only on the shock strength but also
on the details of how the particles are scattered \cite{kirkduffy99}. 
This creates difficulties when studying shock acceleration in flows such as 
spherical, relativistic blastwaves \cite{blandfordmckee76}, the
energetic particle distribution is the solution of a kinetic
equation which describes scattering in the local fluid rest frame.

However, there has been recent progress with ultra-relativistic
shocks, $\Gamma>10$,  for which it has been shown that the index
of the power-law distribution of energetic particles, which
undergo isotropic scattering, is $q=2.23$ where
$N(E)\propto E^{-q}$ is the differential number of particles with
energy $E$ \cite{kirketal00}. This result has potentially important 
implications for fireball models of GRBs (Cavallo \& Rees 1978; 
Goodman 1986; Paczy\'nski 1986) in which a large amount of
energy is suddenly released in the form of an optically thick
electron-positron plasma. The presence of baryonic matter 
\cite{reesmeszaros92} results in the conversion
of the initial energy into a relativistically moving bulk flow.
The forward and reverse shocks will accelerate particles by
the above mechanism in the presence of magnetic turbulence.
Similarly, internal shocks can occur because of variability in the
central engine of a GRB and these are also sites of relativistic
shock acceleration. In this paper we present a model for the
synchrotron emission of shock accelerated particles downstream of
the forward and reverse shocks in a spherically symmetric relativistic
blastwave. In contrast to the strict fireball model our initial
conditions are for a gas at rest with a large amount of internal
thermal energy confined to a small volume. As in the fireball case
the gas will accelerate to the point where all of the initial
energy is converted into bulk kinetic energy if the external
medium is sufficiently tenuous. Ultimately the flow will become
self-similar when the amount of swept up mass becomes significant
and the flow will decelerate thereafter.

Our principle aim in this paper is to study the influence of an evolving,
inhomogeneous flow on the energetic particles produced by the first order Fermi
mechanism at relativistic shocks. This is of central importance in the theory
of GRBs and many papers have recently approached related problems on a number
of different levels. The adiabatic evolution of a GRB interacting with an
external medium, and the associated emission, has been considered
\cite{panaitescukumar00} where an analytic model is introduced for the Lorentz
factor of the shock front and relativistic remnant. A similar hydrodynamical
model has also been employed in Moderski, Sikora \& Bulik 
\shortcite{moderskietal00} for the deceleration of the blast wave but for 
beamed ejecta. Such a geometry has also been discussed in a recent paper 
\cite{granotetal00} where a full hydrodynamical simulation is carried out. 
In this paper we solve the full hydrodynamical equations for a spherically 
symmetric system which starts from rest but begins to expand 
relativistically, and ultimately becomes self-similar, as a result of the 
enormous amount of internal energy in the central engine. Moreover,
particles are accelerated at all shock waves which arise in our simulation,
which in this case are the forward and reverse shocks, with a spectrum
determined by acceleration theory. This model will be generalised to jet type
geometries in future work.

In order to achieve this we have developed a hydrodynamical code, described in
Sect.\ \ref{hydrodynamics} and Appendix \ref{numerical_appendix}, which 
integrates the relativistic Euler equations reliably for Lorentz factors of 
up to several hundred.  This code is applicable to a gas with a general 
equation of state in one dimension and incorporates adaptive hierarchical 
mesh refinement.  In Sect.\ \ref{test} we present tests of our
hydrodynamical code.  Our model for the injection of accelerated particles
into the flow, and their subsequent cooling, is presented in
Sect.\ \ref{synchrotron}.  The initial conditions are
presented in Sect.\ \ref{initial}, while in Sect.\ \ref{hydro_disc} we 
give a discussion of the hydrodynamics of blastwaves with non-zero initial 
radius. The results are presented and discussed in
Sect.\ \ref{results}.

\section{Hydrodynamics}
\label{hydrodynamics}
We wish to simulate a spherically symmetric explosion which gives rise
to relativistic velocities.  We use the relativistic Euler system of equations 
in order to calculate the evolution of this system.  The code used is a
second-order Godunov-type scheme (e.g. van Leer 1977) which uses a linear 
Riemann solver for shocks, and a non-linear one for strong rarefactions.  
Appendix \ref{numerical_appendix} gives the details of the code, along with 
some test calculations.  In this section we briefly describe the equations 
used.

The conservation equations for inviscid relativistic hydrodynamics in 
spherical symmetry are
\begin{eqnarray}
{\partial \over \partial t}\left(\Gamma \rho\right) + {1 \over r^2}
{\partial \over \partial r} \left(r^2 \Gamma \rho \beta\right) & = & 0
\label{cons_den} \\
{\partial \over \partial t} \left(w \Gamma^2 \beta\right) +
{1 \over r^2}{\partial \over \partial r} \left[r^2
\left(w \Gamma^2 \beta^2 + p\right)\right] & = & {2 p \over r}
\label{cons_mom} \\
{\partial \over \partial t}\left(w \Gamma^2 - p\right) + {1 \over r^2}{\partial
\over \partial r}\left(r^2 w \Gamma^2 \beta \right) & = & 0
\label{cons_energy}
\end{eqnarray}
\noindent where $\Gamma$ is the fluid Lorentz factor, $\rho$ is the
proper density, $\beta$ is the velocity in $c=1$ units, $w$ is the enthalpy
and $p$ is the proper pressure.  Time, $t$, and distance, $r$, refer to the
coordinates measured in the observer's frame.  We can relate the enthalpy,
density and pressure by
\begin{equation}
w = \rho + {\gamma_* \over \gamma_* - 1} p
\end{equation}
If the gas is composed of $N$ species which each have mass $m_i$ and
number density $n_i$ then, for a Synge gas \cite{fallekomissarov96},
\begin{equation}
{\gamma_* \over \gamma_* - 1} = \left[\sum_{i=1}^N m_i n_i
\left({K_3(\xi_i) \over K_2(\xi_i)}-1\right)\right]
\left[\sum_{i=1}^N {m_i n_i \over \xi_i}\right]^{-1}
\end{equation}
where $\xi_i={m_i \over k T}$ and $K_\alpha$ are the modified
Bessel functions.  In this work we assume two species - electrons and
protons - which are in thermal equilibrium.  The variable $\gamma_*$ defined 
above is
different to the ratio of specific heats, $\gamma$ which is used in the
calculation of, for example, sound-speeds.  With this relation we should
be able to solve the evolution equations \ref{cons_den} to \ref{cons_energy}.
However, it should be noted that in relativistic hydrodynamics there are
no explicit relations giving the primitive variables $\rho$, $\beta$ and
$p$ in terms of the conserved variables $\Gamma \rho$, $w \Gamma^2
\beta$ and $w\Gamma^2 -p$.  Therefore an iterative algorithm must be
developed to do this.  The precise nature of this algorithm can
affect how high a Lorentz factor can be simulated by the code.  We have
found that the code used here will simulate flows up to Lorentz factors
of at least several hundred reliably.

\section{Particle acceleration and synchrotron emission}

\label{synchrotron}

\subsection{The particle distribution}

Relativistic shocks, in the presence of magnetic fluctuations which
enable multiple shock crossings, can accelerate particles by the 
diffusive shock mechanism \cite{kirkschneider89}.  As described in the 
introduction it has been shown that at 
ultra-relativistic shocks this produces a power law distribution of
particles with index $2.23$, i.e. $N(E)\propto E^{-2.23}$, so that
the energy density does not diverge at high particle energies.
The
acceleration timescale in relativistic flows will scale roughly with the
particle mean free path in the turbulent magnetic field divided by the shock
speed; although there is no published derivation of the precise acceleration
timescale. In this paper we are interested in what happens to the power law
population produced by highly relativistic shocks over the much longer
timescale which is the size of the blastwave divided by $c$. Therefore, over
hydrodynamical timescales, we can treat acceleration as an impulsive injection
of energetic particles, with this universal power law up to arbitrarily high
energies, into each fluid element which passes through a relativistic shock.
We will address the role of an intrinsic cut-off in the shock accelerated
population, either as a result of a finite acceleration timescale or loss
processes, in a future paper. The particles subsequently
suffer synchrotron and adiabatic losses from which we can
calculate the emission. We need to introduce three free parameters
which, when taken together with the results of our hydrodynamical
simulations, will give a complete picture of the blastwave's
evolution and the associated emission of energetic particles.
These parameters are

\begin{itemize}

\item the ratio, $\epsilon_b$, between the magnetic field energy density
and the thermal energy density,

\item the fraction, $\epsilon_e$, of the downstream thermal energy
density which is converted into high energy electrons as a fluid
cell is shocked and,

\item the energy, $E_{\rm min}$, of the lowest energy particle
which is produced at the shock. The differential energy spectrum
is a power law in energy of index $2.23$ from $E_{\rm min}$.

\end{itemize}

Once an electron is injected at the shock it will be scattered in
the local fluid frame and suffer both synchrotron and adiabatic
losses. If the length scale over which the electrons are
isotropised is much shorter than any other length scale of
interest then the relativistic electrons will respond
adiabatically to the expansion or contraction of the flow.
Adiabatic losses, or indeed gains, are then described by the fact
that $p/\rho^{1/3}$ is constant where $p$ and $\rho$ are the
particle momentum and fluid density in the local fluid frame. With
$E=pc$ for ultra-relativistic particles the combined synchrotron
and adiabatic losses are described in the comoving frame by

\eqb \dot{E}=-\alpha B^2E^2+{1\over 3}{\dot\rho\over\rho}E
\label{energyloss} \eqe

\noindent where $\alpha$ is a constant and $B$ the magnetic field strength
in the local fluid frame. 
Consider now a fluid element which is
shocked at time $\that_0$ when a power law distribution of
energetic particles is injected and where $\that$ is time measured
in the comoving frame. This population then evolves according to
the equation

\eqb {\partial N\over\partial\that}+{\partial\over \partial
E}\left(\dot{E}N\right)=Q(E,\that) \label{kineticeqn}\eqe

\noindent where $N(E,\that)$ is the differential number of particles of
energy $E$ at time $\that$. The losses, $\dot{E}$, are given by
equation (\ref{energyloss}) while the injection term is
$Q(E,\that)=Q_0\delta(\that-\that_0)H(E-E_{\rm min})E^{-p}$ ($H$ being
the Heviside function) which describes injection of a power law spectrum, 
starting at time $\that_0$ with a minimum particle energy $E_{\rm min}$. 
We can solve equation (\ref{kineticeqn}) by finding the characteristic 
curves along which $N(E,\that)dE=N(E_0,\that_0)dE_0$ where a particle with
energy $E_0$ at $\that_0$ cools to an energy $E$ at time $\that$. From 
equation (\ref{energyloss}) we have

\eqb {d\over d\that}\left(E^{-1}\rho^{1\over 3}\right)
=\alpha\rho^{1\over 3}B^2\eqe

\noindent which can be solved to show that along a characteristic curve

\eqb E_0={\rho_0^{1\over 3}E\over \rho^{1\over 3}-\alpha
E\int_{\that_0}^{\that}\rho^{1\over 3}B^2
d\that}\label{charcurve}\eqe

\noindent $N$ is conserved so that

\eqb {dE_0\over dE}={\rho_0^{1\over 3}\rho^{1\over 3}\over
\left(\rho^{1\over 3}-\alpha E\int_{\that_0}^{\that}\rho^{1\over
3}B^2d\that\right)^2}\eqe

\noindent and the solution to (\ref{kineticeqn}) becomes

\eqb
\label{kinetic_sln}
N(E,\that)=Q_0\rho_0^{1-p\over 3}\rho^{1\over
3}E^{-p}\left(\rho^{1\over 3}-\alpha
E\int_{\that_0}^{\that}\rho^{1\over 3}B^2d\that\right)^{p-2}
\eqe

In order to calculate the particle spectrum we need to evaluate the 
integral in equation (\ref{kinetic_sln}) for each fluid element which 
has passed through a shock at some point. No fluid element passes through 
two or more shocks during these simulations; our method would have to be 
adapted in such a case (e.g.\ for overtaking internal shocks).
We introduce a continuity equation for the integral quantity in 
equation (\ref{kinetic_sln}). Define $I \equiv \int_{t_0}^{t_0+t} {\rho^{1 \over 3} B^2
\over \Gamma} dt$ where the factor of $\Gamma^{-1}$ boosts observer time to the comoving fluid time.  Then the equation for the distribution of $I$ is simply
\begin{equation}
{\partial \over \partial t}\left(\Gamma \rho I\right) + {1 \over r^2}
{\partial \over \partial r} \left(r^2 \Gamma \rho I \beta\right) = B^2
\rho^{4 \over 3}
\end{equation}
Thus the quantity $I$ increases by $\rho^{1/3}B^2\Delta t/\Gamma$ with each
time-step $\Delta t$, following the fluid element, as required.  If the 
gradient of $\Gamma\beta$ is ever less than a certain negative value, chosen 
to ensure that the fluid element has passed through a shock, then $I$ 
is set to 0 at that point. The above conservation equation can easily be 
solved in conjunction with the hydrodynamical equations using the same 
numerical methods.  Note that $I$ is not kept at zero in unshocked
fluid, but this does not affect the synchrotron emission since, in such
fluid, there are no energetic particles.

\subsection{Synchrotron emission}

We approximate the emission of a single particle to be a delta function in
frequency with a single particle emissivity in the fluid frame given by 
$j_{\nuf}(\gamma)=a_0\gamma^2B^2\delta(\nuf-a_1\gamma^2B)$ where 
$\gamma$ is the Lorentz factor of the electron in the fluid frame, 
$a_0=\sigma_Tc/6\pi$ and $a_1=e/2\pi m_ec$ with $\sigma_T$ the Thomson cross 
section. With the local electron spectrum given by \ref{kinetic_sln}, 
the emissivity in the local fluid frame is then

\begin{equation}
\emf=\int j_{\nuf}(\gamma)N(\gamma,\that)\,d\gamma
\end{equation}

\begin{figure}
\begin{center}
\leavevmode
\epsfxsize = 240pt \epsfbox{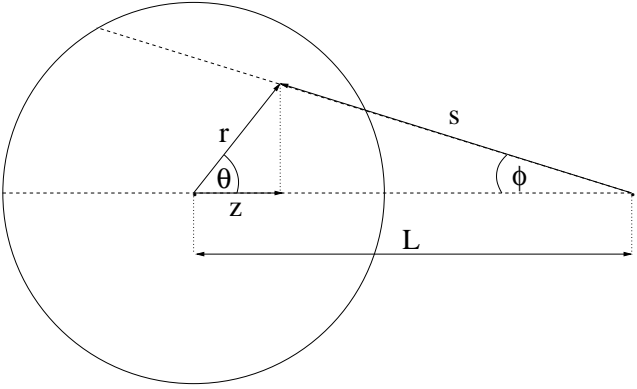}
\caption{\label{synch_fig} Diagram of the technique used to calculate the
spectrum of photons arriving at an observer at a given time $t(obs)$.  The
fluid element is located at $r$ in the simulation, and the current time is
$t$.  This, in conjunction with $L$, the distance between the observer
and the blastwave, defines $\theta$ and $\phi$.  See text.}
\end{center}
\end{figure}

The emissivity in the observer's frame is related to that in the fluid frame
by $\emo=D^2\emf$. The Doppler factor, $D$, for the fluid element with a three
velocity $\beta$ making an angle $\theta$ to the line of sight is 
$D=[\Gamma(1-\mu\beta)]^{-1}$ where $\mu=\cos\theta$. The observed frequency
is related to the photon frequency in the fluid frame by $\nu=D\nuf$. The 
observed intensity of radiation at an inclination $\phi$ to the line between
the observer and the centre of the explosion is
\eqb
I_\nu(\phi,t_0)=\int_0^\infty\emo(s,\phi,t)\,ds
\eqe
where $t_0$ is the time of observation and $t=t_0-s/c$ is the time of emission
in the observer's frame a distance $s$ away from the observer (figure 
\ref{synch_fig}). The observed flux density is obtained by 
integrating the intensity over the entire source
\eqb
F_\nu(t_0)=\int_\Omega I_\nu(\phi,t_0)\,d\Omega
\eqe
In cylindrical coordinates $(z,r)$ centred on the explosion, $s\approx L-z$
since $\phi\ll 1$ so that 
\eqb
F_\nu(t_0)={2\pi\over L^2}\int_{-r_c}^{r_c}dz\int_0^{r_c}
\emo(z,r,t)r\,dr 
\eqe
with $L$ the distance from the observer to the centre of the explosion and
$r_c$ the size of the computational domain. If $r_0$ is the initial radius of
the explosion then the interval of $t_0$ since the arrival of the first 
photon is $\overline{t_0}=t_0-(L-r_0)/c$ so that the flux can be written as
$F_\nu(\overline{t_o})$. By the change of variable 
$z=r_0+c(t-\overline{t_0})$ the integration over $z$ can be converted to 
one over $t$, the time of emission in the observer's frame,
\eqb
F_\nu(\overline{t_0})={2\pi c\over L^2}\int_0^{t_{\rm end}}dt
\int_0^{r_c}\emo(r_0+c(t-\overline{t_0}),r,t)r\,dr
\eqe

In the simple case where the fluid is assumed to move directly towards the 
observer with a constant velocity we have successfully compared our method 
with the exact results. The above analysis, which is essentially the same as
in Komissarov \& Falle \shortcite{komissarovfalle97}, also corresponds for 
the latter simple case to the treatment used in Moderski et al.\ 
\shortcite{moderskietal00}. In particular from those particles which have 
not radiated away a significant fraction of their energy we get the 
{\it uncooled} spectrum of $F_\nu\propto\nu^{-(p-1)/2}$ while the higher 
energy particles have a steeper, {\it cooled} spectrum of 
$F_\nu\propto\nu^{-p/2}$. Below the minimum observed frequency the flux 
scales as $F_\nu\propto\nu^{1/3}$.  At even lower frequencies the effect of 
synchrotron self-absorption will become important, although we have not 
explicitly calculated this, leading to a $\nu^2$ part of the spectrum.

\section{Initial conditions}

\label{initial}

In studies of the fireball model of gamma-ray bursts, the parameters
used are the total mass, $M$, and the initial radius, $R_0$.  In general, 
$E$, the total energy, is thought to lie somewhere in the region 
$10^{51}$-$10^{54}$ ergs.  The ratio between the rest-mass
energy in the blast (i.e.\ the mass of the baryonic component), and the
total energy determines two critical radii.  These are the radius at
which the baryons have been accelerated up to their maximum velocity,
$R_c$, and the radius at which the shell of ejected baryons have swept
up their own mass in interstellar material, $R_d$.  This latter radius
is the radius at which the baryonic shell begins to decelerate.  The
former also depends on $R_0$, the initial radius of the blast.  This
initial radius is thought to be quite small, with $R_0 \sim 10^{12}$ cm.

In order to get all the initial energy in the blast converted into
kinetic energy of the baryons, it is also required that the fireball be
optically thick to pair creation until the radius of the blast exceeds
$R_c$.  If this were not the case, then the photons would escape into
space before accelerating the baryons to their maximum velocity.
In our case we put the initial energy of the blast into thermal
pressure.  This means that it is already in the kinetic energy of the
baryons.  Hence the latter consideration need not be taken into account
in our initial conditions.

The results of two simulations are presented in this paper.  Each of the
simulations used the following properties:
\begin{itemize}
\item $E=10^{51}$ ergs
\item Ratio of energy to mass: $\eta = {E \over M c^2} = 580$.  This
value being chosen so that the Newtonian value of $R_d$ is twice $R_c$.
In reality, $R_d$ is slightly less than $R_c$ due to relativistic
effects.
\item $R_0 = 1.2\times10^{14}$ cm.  This is much larger than the
postulated physical value.  However, reducing $R_0$ to $10^{12}$ cm
would impose extremely severe computational costs on the simulation.
This increase in the value of $R_0$ is not expected to change the
properties of the synchrotron emission calculated as we are interested
in the afterglow in this work, and not the emission from the
acceleration phase.
\item $\epsilon_e = 0.01$ - this value is consistent with observed
values found by, e.g.\ Wijers \& Galama \shortcite{wijersgalama99}
\item $\epsilon_b = 0.01$ in one simulation, and $\epsilon_b=0.1$ in the
other.  This is the only way in which the two simulations differ.
\end{itemize}

We must also choose the injection energy $E_{\rm min}$ for the shock
acceleration process. Physically this will be set by nonlinear plasma
processes. The highest energy {\it thermal} particles downstream of a shock
can leak into the upstream forming a beam of particles which can excite plasma
turbulence. This turbulence then scatters particles back into the downstream
fluid and multiple shock crossings become possible. The problem of
determining the details of this process is an unsolved one in the theory of 
particle acceleration at relativistic shocks.  However, it is clear that the 
value $E_{\rm min}$ must correspond to the high energy
tail of the downstream thermal population. This is hardly surprising since the
thickness of a shock wave will be set by the gyroradius of a thermal particle
in a collisionless plasma and first order Fermi acceleration will only be
applicable to particles with gyroradii much larger than this. Consequently we
have fixed the value of $E_{\rm min}=10m_ec^2$ which is chosen so that
it exceeds the {\it thermal} energy for all shocked fluid elements. Setting
$E_{\rm min}$ as a factor of a few times the thermal energy at the instant a
piece of fluid is shocked does not change our results appreciably.
 
Initially, then, there is a sphere of radius $R_0$ which is at rest at
the origin.  This sphere contains all the initial energy and mass of the
blast.  Outside this sphere is a gas with a density of 1 cm$^{-3}$.

\section{Hydrodynamic evolution of a spherical blastwave}

\label{hydro_disc}

\begin{figure}
\begin{center}
\leavevmode
\epsfxsize = 240pt \epsfbox{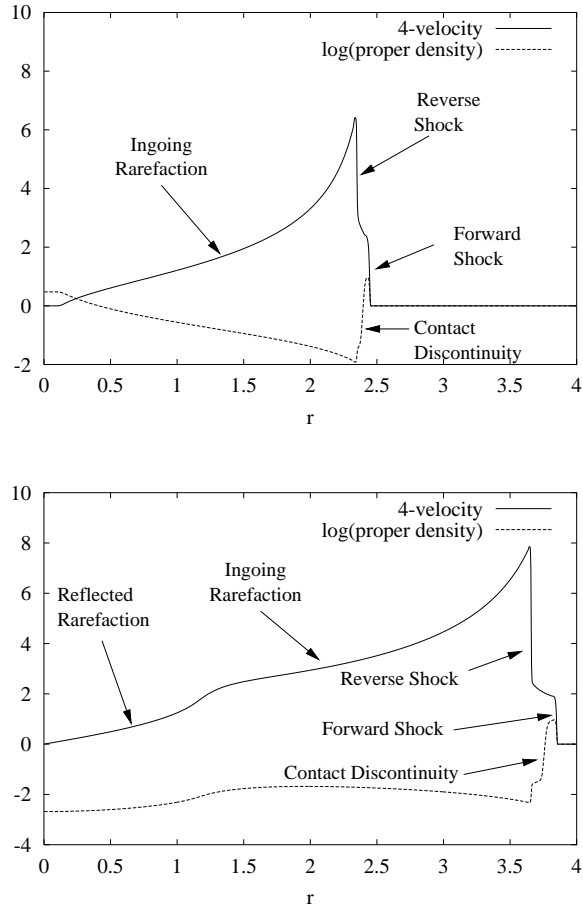}
\caption{\label{early_times}  Plots of the spatial component of the
4-velocity, along with the proper density for times of $t=1.5$ (top) and
$t=3$ (bottom).  All the waves in the system are labelled.  See text.}
\end{center}
\end{figure}

In this section we discuss the qualitative evolution of a spherical
blastwave with finite initial radius.  This discussion applies to both
relativistic and non-relativistic cases.  The purpose is to elucidate,
in a qualitative fashion, how the forward and the reverse shocks are 
formed, and how they behave, from a hydrodynamic perspective.  Figure
\ref{early_times} shows the early-time evolution of the system, computed
using the code presented here.  The initial conditions used were as
follows:

\begin{eqnarray*}
\rho & = & \left\{\begin{array}{ll}
                    3 & \mbox{on $r\leq1$} \\
                    1 & \mbox{on $r > 1$}
                    \end{array}
            \right. \\
u & = & 0 \\
p & = & \left\{\begin{array}{ll}
                    10^3 & \mbox{on $r\leq1$} \\
                    3\times10^{-3} & \mbox{on $r > 1$}
                    \end{array}
            \right.
\end{eqnarray*}

These initial conditions are similar to those in Sect.\
\ref{code_test_3}, and are chosen so that the relevant features are
clearly visible in the results.

Initially a shock is driven into the ambient medium (the ``forward
shock'' in figure \ref{early_times}), while a rarefaction propagates from 
$r=R_0$ back to $r=0$ (the ``ingoing rarefaction'' in figure 
\ref{early_times}).  Note that, at $t=1.5$, the ingoing rarefaction has 
still not reached the origin.  The forward shock gets progressively stronger 
as blast material gets accelerated down the pressure gradient of the 
rarefaction.  There is also an entropy wave (the ``contact discontinuity'') 
which moves out from $R_0$ with a speed slightly lower than that of the 
blastwave.

In addition, a reverse shock is created {\em almost immediately} between
the entropy wave and the rarefaction (see figure \ref{early_times}).  If the 
system were in planar symmetry then one would not expect this reverse shock 
since there are only three characteristics in hydrodynamics (see, 
e.g.\, Landau \& Lifschitz 1966).  We get this fourth 
wave purely as a result 
of the non-planar nature of the system.  It forms because, as the forward 
shock moves outwards it sweeps up an increasing amount of interstellar matter 
per unit distance.  This means that the forward shock does not move outwards 
as fast as it would in the planar case, and the reverse shock is formed to 
decelerate material behind the forward shock to the appropriate speed.  

When the rarefaction reaches $r=0$ it is reflected and enhanced.  This
``reflected rarefaction'' can be seen clearly at $t=3$ in figure 
\ref{early_times}.  This rarefaction will eventually catch up with the 
reverse shock/blastwave system (see figure \ref{second_hydro} and Sect.\ 
\ref{hydro_results}).  When the rarefaction catches up with the reverse 
shock, the ram-pressure of the material entering this shock is drastically 
reduced, because it has been rarefied and decelerated.  Hence it can no 
longer support the thermal pressure between the forward and reverse shocks.  
As a result, the material between these two shocks expands.  Since the 
ram-pressure of the material entering the forward shock has not changed 
(assuming constant external density), this necessitates the reverse shock 
slowing down with respect to the rest-frame of the initial blast.  Indeed, it 
will actually begin propagating towards the origin.  The speed at which it
does this depends on the initial value of $\eta$, since this determines
the strength of the initial (and so the reflected) rarefaction.  For
example, if $\eta$ is very small, then the force due to the pressure
gradient will produce a relatively small increase in velocity of the
material in the initial sphere.  Hence the rarefaction will be weak.
If, on the other hand, $\eta$ is large then the force due to the
pressure gradient will accelerate the blast material to high velocities
very quickly, producing a very strong rarefaction.  For the
conditions in Sect.\ \ref{initial} it becomes sufficiently strong to
cause ejected material to flow back towards the origin also.

When the reverse shock reaches $r=0$ it rebounds and, as it propagates
outwards again, it weakens.  Once this shock is sufficiently weak to be
ignored, we have reached the self-similar stage of evolution.  Prior to
this point the finite radius of the initial blast plays a part, and
hence there is a significant length-scale in the system.  If the blast
was initially of zero radius then the whole process described above
would happen infinitely quickly (or, equivalently, would not happen at
all).

We can relate the above ``hydrodynamical'' picture with the one used by,
for example, Rees \& M\'esz\'aros \shortcite{reesmeszaros92}, as follows. 
The 
 coasting radius, $R_c$, is reached when all material has been
accelerated 
 down the pressure gradient of the ingoing rarefaction.  This, of
course, 
 never occurs, but one can define $R_c$ as being the radius when
``most'' of 
 the material has been accelerated.

The deceleration radius, $R_d$, is then the radius at which the reflected
rarefaction catches up with the reverse shock.  This is when the
material between the forward and reverse shock expands back towards the
origin, necessarily decelerating significantly as it does so.

\section{Results}
\label{results}

We discuss the results of the simulations in two sections.  The first
deals with the detailed hydrodynamic evolution of the blastwave.  The
second concerns the observed spectra and light-curves.

\subsection{Hydrodynamic results}
\label{hydro_results}

\begin{figure}
\begin{center}
\leavevmode
\epsfxsize = 240pt \epsfbox{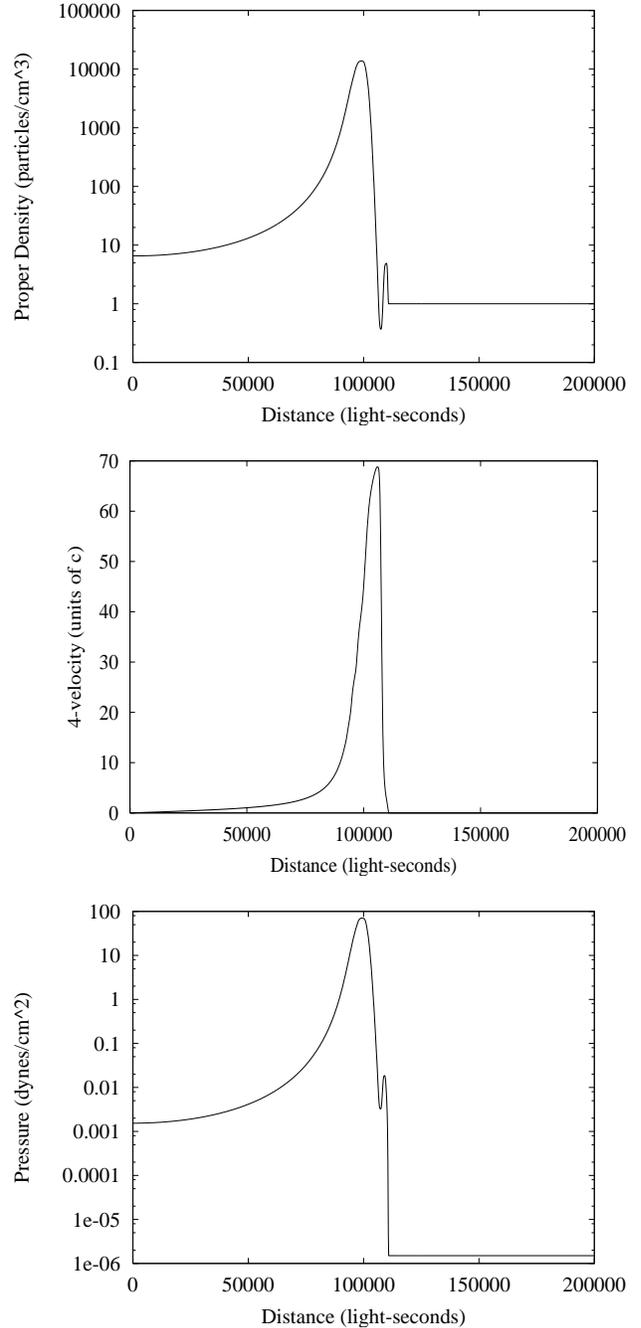}
\caption{\label{first_hydro}  Plots of density, 4-velocity and pressure
(top to bottom) for a time of $1\times10^5$ seconds after the initial
blast.  We can see the original and reflected rarefaction, as well as
the forward and reverse shocks.  See text.}

\end{center}
\end{figure}

Here we show the hydrodynamic aspect of the results of our simulations.
Figure \ref{first_hydro} shows the distribution of proper density,
velocity and pressure after $1\times10^5$ seconds.  We can see the two
rarefactions mentioned in Sect.\ \ref{hydro_disc}.  The head of the
reflected rarefaction lies at the location of the peak velocity, with
its tail at $r=0$.  The original rarefaction lies between the peak of
velocity and the rise in the density and pressure plots at
$\sim 1\times10^5$ light-seconds.

The forward shock can be identified as the right-most rise in pressure
and density, while the reverse shock is just to the left of this.

\begin{figure}
\begin{center}
\leavevmode
\epsfxsize = 240pt \epsfbox{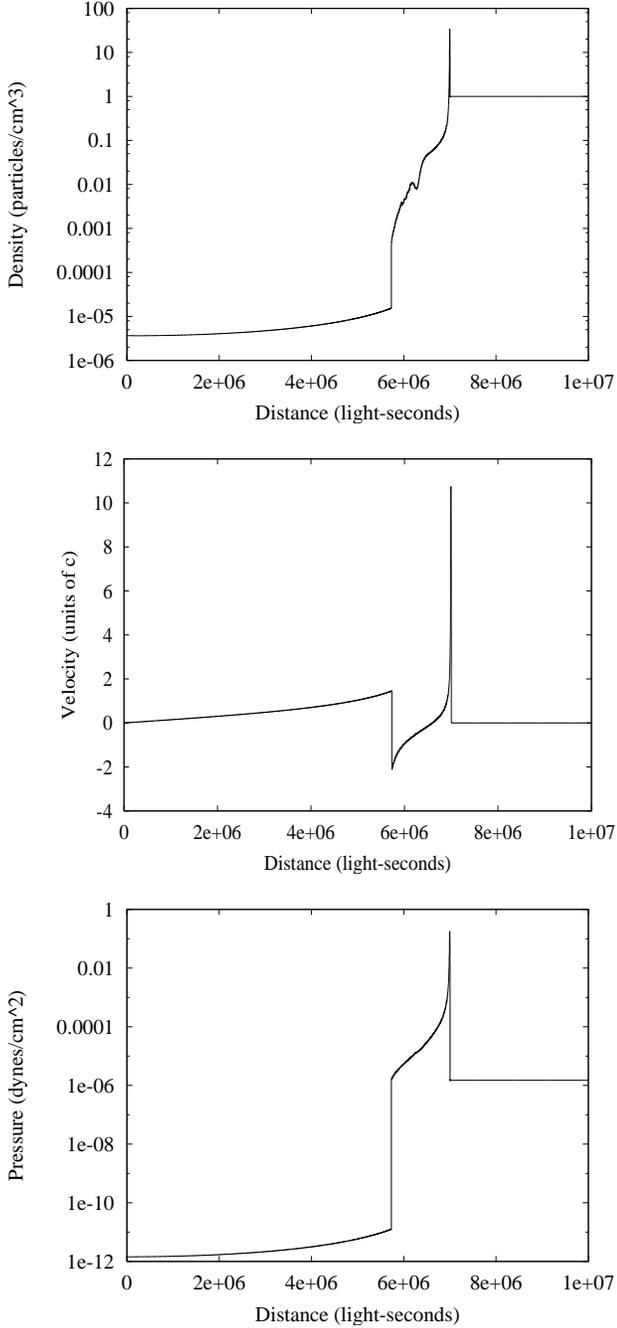}
\caption{\label{second_hydro}  As for figure \ref{first_hydro}, but for
a time of $7\times10^6$ seconds after the initial blast.}

\end{center}
\end{figure}

Figure \ref{second_hydro} shows the same plots as figure
\ref{first_hydro}, but for a time of $7 \times 10^6$ seconds.  Here we
can see the reverse shock has begun to separate significantly from the
forward shock.  The reflected rarefaction is now the only rarefaction
present in the system.  The reverse shock is expanding back into this
rarefaction due to the insufficient ram-pressure of the material in the
reflected rarefaction to support the pressure of the material between
the forward and reverse shocks.

It is possible to see oscillations in the density plot just behind the
reverse shock.  These oscillations arise out of numerical errors during
the time that the reflected rarefaction comes close to, but is not in
contact with, the reverse shock.  The rarefaction is extremely strong
here, and it is not surprising that numerical errors become significant.
However, the influence of these errors on the overall solution, or on
the calculated synchrotron emission, is negligible due to the low density
of the region containing the errors.

\begin{figure}
\begin{center}
\leavevmode
\epsfxsize = 240pt \epsfbox{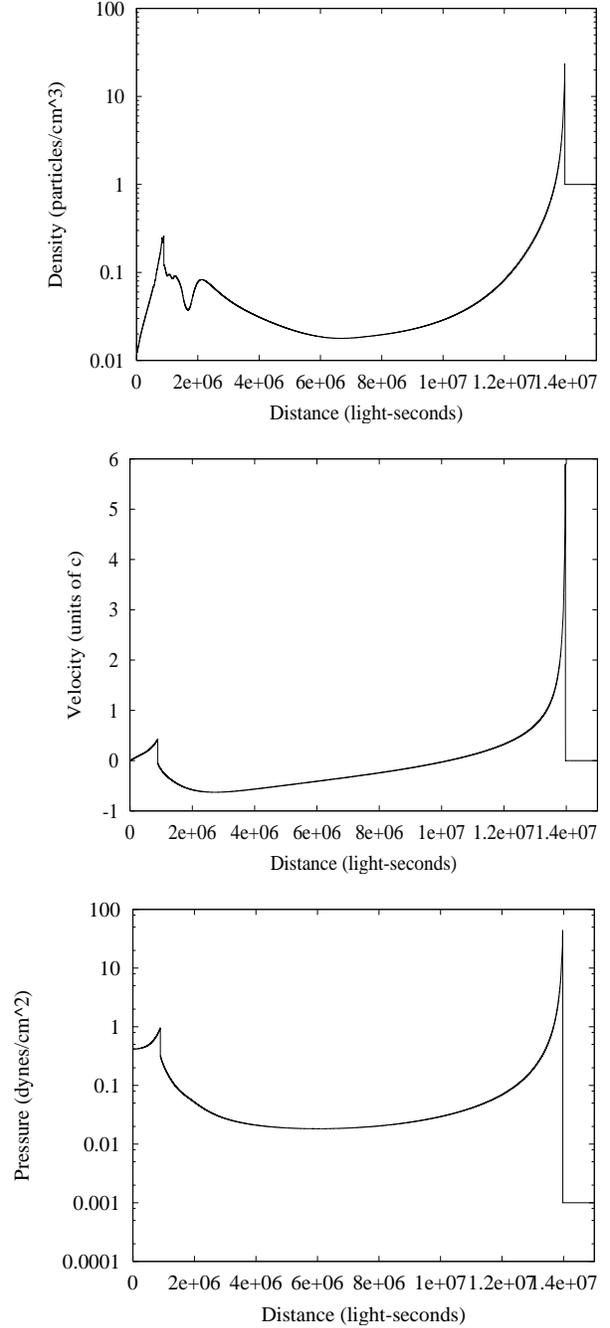}
\caption{\label{third_hydro}  As for figure \ref{first_hydro}, but for a
time of $1.4\times10^7$ seconds after the initial blast.}

\end{center}
\end{figure}

Figure \ref{third_hydro} shows the latter stages of the non-self-similar
evolution of the system.  At this stage the reverse shock has propagated
all the way to $r=0$ and has rebounded.  We can see that it has weakened
greatly, and will continue to do so.  Once this shock is sufficiently
weak to be ignored the system will evolve in a self-similar way.  We can
see the entropy errors generated by the very strong reflected
rarefaction remaining just ahead of the reverse shock now.  These
errors, although apparently significant here, do not affect the
synchrotron results due to the low densities in this region, and the low
relativistic boosting suffered by emission from this material.  Both
these effects make the influence of emission from this material
negligible.

\subsection{Spectra and light curves}

In this section we present the results from the calculations of the
synchrotron emission.  

\subsubsection{Spectra}

\begin{figure}
\begin{center}
\leavevmode
\epsfxsize = 240pt \epsfbox{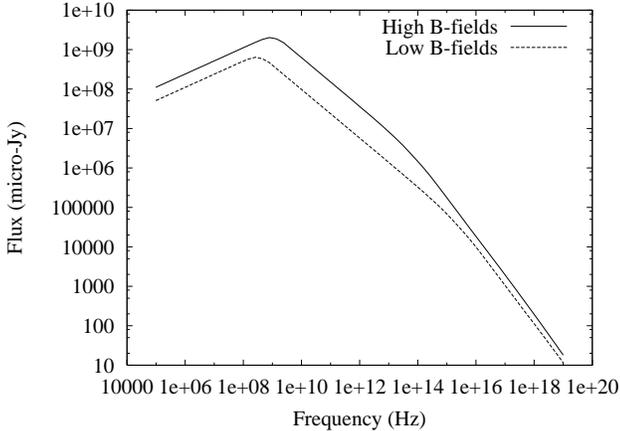}
\caption{\label{spectrum}  The spectrum observed at 24 hrs after the
initial blast observation for the case of $\epsilon_b=0.01$ (``Low
B-fields'') and $\epsilon_b=0.1$ (``High B-fields'').}
\end{center}
\end{figure}

Figure \ref{spectrum} shows a plot of the spectrum observed 24 hours after 
the initial blast could have been observed for the case where 
$\epsilon_b=0.1$ (hereafter referred to as the high $\epsilon_b$ case) and 
for $\epsilon_b=0.01$ (hereafter referred to as the low $\epsilon_b$ 
case).  It is clear that in both cases the spectrum is a broken power-law.  

In the first section of the spectra, the flux goes as $\nu^{1/3}$, as
expected for synchrotron radiation below the peak frequency of the lowest
energy electron. At lower frequencies, which are not plotted here, synchrotron
self-absorption would give a flux of $\nu^{1/3}$. The second part of the
spectrum is, in both cases, the power law $\nu^{-0.615}$ indicating that this
part of the spectrum is dominated by emission from electrons, with a
distribution of $E^{-2.23}$, which have not suffered significant adiabatic 
or synchrotron cooling.  

The break from this part of the spectrum to the final, steeper, part
occurs in different places in the spectra plotted.  For the high
$\epsilon_b$ case, the break occurs at a lower frequency than for the low 
$\epsilon_b$ case as would be expected since, in higher
magnetic fields, the losses suffered by the energetic population are 
correspondingly higher. However, while in the final part of the spectrum 
the exponent predicted from simple theory would be $-p/2=-1.115$, we find the 
spectrum to be slightly harder.  In the high $\epsilon_b$ case, we have 
$F_\nu \propto \nu^{-1.019}$, and in the low $\epsilon_b$ case, 
$F_\nu \propto \nu^{-0.981}$.  The reason for the slightly harder spectra 
at high frequencies is the non-uniform velocity distribution in the
``shell'' of ejected material (see, e.g.,\ figure \ref{second_hydro}).
Material moving at high velocity towards the observer will be more
heavily weighted in the spectra than material moving at lower velocity,
due to relativistic beaming.  Since such high velocity material occurs
immediately behind the forward shock, this material will contribute more
to the spectrum than material further back from the shock.  The fluid
just behind the shock contains an electron population which has only
recently been accelerated and so emission from here will be dominated by
uncooled electrons to a very high frequency.  Material from further
behind the shock will have emission dominated by cooled electrons down
to a lower frequency.  Hence, above a certain cut-off, we expect to get
a mixture between emission from cooled and un-cooled electrons, with
un-cooled electrons being preferentially weighted.  This leads to an
exponent for the spectrum lying between the uncooled value (${1-p \over
2} = -0.615$) and the cooled value (${-p \over 2} = -1.115$).
This conclusion is further supported by the fact that the high
$\epsilon_b$ case gives a slightly steeper spectrum at these frequencies than 
the low $\epsilon_b$ case.  For high magnetic fields the contribution from 
cooled electrons will be stronger closer to the shock than in the low 
magnetic field case.  Therefore, while we still expect the uncooled
electrons to be preferentially boosted in the spectrum, the effect of
the cooled electrons will be stronger for higher magnetic fields.

It is worth discussing the behaviour of the `critical' frequencies in
the spectra with time - i.e.\ where the breaks occur.  The lower break, which
arises due to the presence of electrons injected with the minimum energy
$E_{\rm min}$ behaves roughly as $t^{-0.73}$.  This is the same for both
the low and high $\epsilon_b$ cases.  This is a much less
dramatic decrease with time than that predicted by Sari \& Piran
\shortcite{saripiran97} and is due to the fact that, here, we define the
minimum energy of injection {\em a priori}.  This energy is chosen so
that the gyroradius of an electron at this energy would be accelerated
by the Fermi mechanism.  The latter authors choose to define the {\em
number of electrons} accelerated, and this, in conjunction with the
energy in energetic particles, defines $E_{\rm min}$.  This produces the
different behaviours of the lower break in the spectra with time.

We find that the frequency of the second (upper) break goes like
$t^{-1.2}$ for the low $\epsilon_b$ case and $t^{-1.75}$ for the high
$\epsilon_b$ case.  This is a much {\em faster} decrease with time than that
predicted in Sari \& Piran \shortcite{saripiran97}.  This difference cannot
be due to the different choice of definition of $E_{\rm min}$, but it could
be hypothesised that it is due to the effects of adiabatic cooling.  This
cooling will cause the second critical frequency to decrease faster
where the flow is divergent.  However, simulations of this system
without adiabatic cooling show that, while adiabatic cooling does have
this effect on the observed spectrum, it is too small to explain the 
discrepancy.  It is not clear what specifically causes this difference
between our results and those of Sari \& Piran \shortcite{saripiran97},
but it must be remembered that here we perform a much fuller treatment
of the dynamics and so it is not surprising that we get some significant
differences with semi-analytic approaches.

\subsubsection{Light curves}

\begin{figure}
\begin{center}
\leavevmode
\epsfxsize = 240pt \epsfbox{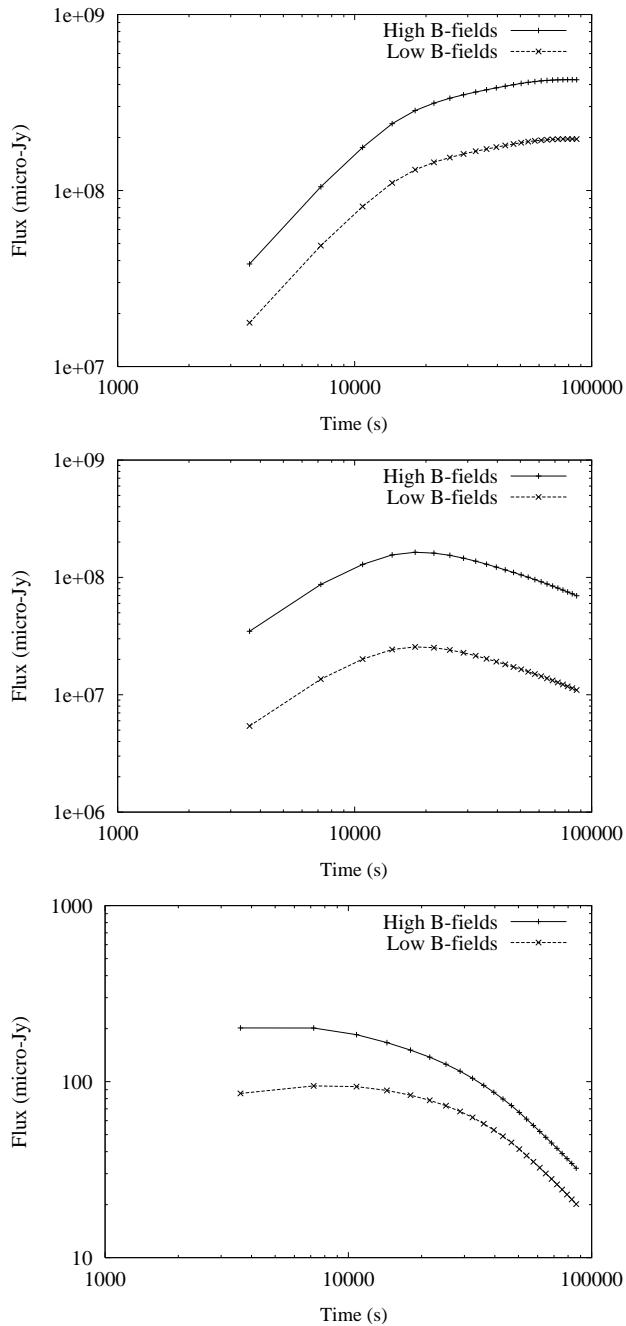}
\caption{\label{lightcurves} Light curves calculated at low,
intermediate, and high frequencies.  See text.}
\end{center}
\end{figure}

In this section we discuss the light curves resulting from the
simulations described above.  Figure \ref{lightcurves} shows plots of
the light curves at low, medium and high frequencies.  These frequencies
are chosen so that the points are always in the regime where $F_\nu
\propto \nu^{1/3}$ (low frequencies), $F_\nu \propto \nu^{1-p \over 2}$
(medium frequencies) and, for high frequencies, the frequency is always
in the third section of the spectrum.

It should be noted in the following discussion that the range of times
covered by these light curves is 1 to 24 hrs after the first signal of
the blast reaches the observer.  Initially we see an increase in all the
light curves shown, with the exception of the high frequency, high
$\epsilon_b$ case, where the emission is roughly constant over the
first couple of hours.  This increase is much faster than that predicted
by previous analytic work.  For example, for the low $\epsilon_b$ case
at low frequencies, the flux goes as roughly $t^{1.4}$ at early times (cf 
Sari \& Piran, 1997, where the maximum increase is $t^{1/2}$).  This effect 
is not due to any problems with numerical 
resolution since the emission in the 1st hour is dominated by material 
between the forward and reverse shocks approximately $6\times10^6$ seconds 
after the initial blast, when the two shocks are well-resolved and 
separated by the code.

The behaviour of the low frequency light curves is qualitatively
different to the other two.  It keeps increasing for the duration
of the simulation.  This is unsurprising as it will only begin to
decrease when the critical frequency $\nu_{\rm min}$ corresponding to
$\Gamma_{\rm min}$ passes through the frequency of the light curve, and
we have chosen this frequency so that this does not happen.  It is very
difficult to say that the light curve behaves anything like a power-law
with breaks over this time-scale based on our results.

The medium frequency light curves initially increase, as already stated,
and then fall off dramatically.  Again, it is difficult to see any point
at which the behaviour of the light curve is a true power-law.  The high
frequency light curves also begin to fall off very steeply, and again,
it is difficult to see any sign of a power-law behaviour in the curves.

The lack of a clear power-law behaviour in the light curves may well be
due to the restricted time-scale over which the curves are calculated
(from 1 hour to 1 day).  However, it is clear that, if there is a broken
power-law behaviour then the breaks are smeared out, and, in addition,
the final fall-off of the light-curve would seem to be much faster than
previously predicted. 

The unexpected behaviour of the low and medium frequency light curves 
is more likely to do with the different treatment of the minimum energy
in the accelerated electron population.  In our case, since we fix
$E_{\rm min}$ {\em a priori}, we allow the available energy to define
the number of energetic electrons.  Previously it has been common to
define the number of energetic electrons and hence the available energy
defines $E_{\rm min}$.  This will lead to quite different behaviours of
the lower break frequency with time.

\section{Conclusions}

We have presented a model for the synchrotron emission of energetic particles
downstream of relativistic, spherical shock waves. The hydrodynamical part of
the problem has been solved numerically with the simpler simulations agreeing
with results published elsewhere. On the other hand the particle acceleration
aspect has been treated as an {\it injection} process with relativistic shocks
leaving a population of energetic particles immediately downstream. The
spectral index is known from semi-analytic work so that we need only specify
the fraction of the downstream thermal energy which is converted into energetic
particles and a lower cut-off energy. The particles subsequently lose energy
by synchrotron cooling and adiabatic losses. The hydrodynamical results have
captured the evolution of both the forward and reverse shocks, which are of
principal interest for particle acceleration and radiative emission, as well as
the rarefaction waves. The spectra emitted from our system largely agree with
the simple predictions for the low energy, uncooled part of the
spectrum. However, at the higher frequency, cooled, part of the electron 
population the relativistic boosting of the material coming straight 
towards the observer hardens the spectrum from the pure cooled value which 
comes from material further downstream and from material which is not 
moving directly towards the observer. Further, non-trivial behaviour has 
been found for the temporal variation of both the break frequencies and the 
light curves. Adiabatic losses coupled with integrating the spectrum over 
a spherical system tend to smear out the break frequencies which are 
predicted from simple scaling arguments. 

It is clear that we need to include several other effects before bringing our
results into contact with observations. A generalisation to a 2-D
hydrodynamical code has already begun. The role of {\it internal} shocks will
also be included along with their effect on acceleration. The energetic
particles obey a transport equation, the solution of which determines the
spectral index and we are working towards including such a solution in our
model. More complicated radiative processes can also be included 
such as the inverse-Compton and synchrotron self-Compton processes. 
However, what we have done in this
paper is to combine a realistic hydrodynamical model with results from particle
acceleration theory in a relativistic flow. The computed hydrodynamical
evolution, spectra and light curves will allow us to compare this model with
observations in future work.

\section*{Acknowledgments}

We would like to thank Luke Drury for useful discussions on the
incorporation of adiabatic losses into the electron distribution.
We are grateful to an anonymous referee for comments which led to improvements
in the paper. This work was supported by the TMR programme of the European
Union under
 contract FMRX-CT98-0168.

\appendix

\section{Numerical Details}

\label{numerical_appendix}

In this appendix we discuss the details of the code used for the
calculations presented in this paper.  We also present the results of
three test calculations.

We employ a second order finite volume Godunov-type scheme to solve the
equations \ref{cons_den} to \ref{cons_energy} in spherical symmetry.  
Assuming that the grid cells are defined so that cell $i$ occupies the 
space $[r_{i-1/2}, r_{i+1/2}]$ then the scheme can be written as
\begin{eqnarray}
\vec{U}^{n+1}_i = \vec{U}^n_i - {3 \Delta t \over r_{i+1/2}^3 - r_{i-1/2}^3}
\left[r_{i+1/2}^2 \vec{F}_{i+1/2}^{n+1/2} \right. \nonumber \\
\left. - r_{i-1/2}^2 \vec{F}_{i-1/2}^{n+1/2}\right] + \vec{S}_i^{n+1/2} \Delta t
\end{eqnarray}
where superscripts refer to the time index and subscripts refer to the
spatial index.  Also,
\begin{eqnarray}
\vec{U}^n_i & = & \left[\left(\Gamma \rho, \Gamma \rho \beta, w \Gamma^2
- p\right)^T\right]^n_i \\
\vec{F}^n_{i\pm 1/2} & = & \left[ \left( \Gamma \rho \beta, w \Gamma^2
\beta^2 + p, w \Gamma^2 \beta\right)^T\right]^N_{i \pm 1/2} \\
\vec{S}^n_i & = & \left[ \left( 0, {2 p \over r}, 0 \right)^T\right]^n_i
\end{eqnarray}
\noindent  It should be noted that the source term $\vec{S}$ should be
volume-averaged over $[r_{i-1/2}, r_{i+ 1/2}]$ \cite{fallekomissarov96}.
So, for spherical symmetry, and using a first order scheme we get that
the source term for the radial momentum equation is
\begin{equation}
S^n_i = 3 p^n_i {r_{i+1/2} + r_{i-1/2} \over r^2_{i+1/2} +
r_{i+1/2}r_{i-1/2} + r^2_{i-1/2}}
\end{equation}
For a second order scheme, which allows $p$ to vary linearly in a
given cell the correct form for $S^n_i$ is 
\begin{eqnarray}
S^n_i & = & 3 p^n_i {r_{i+1/2} + r_{i-1/2} \over r^2_{i+1/2} +
r_{i+1/2}r_{i-1/2} + r^2_{i-1/2}} \nonumber \\
&& +{g_p}^n_i\left[ {9 (r_{i+1/2} + r_{i-1/2})^2(r^2_{i+1/2} + r^2_{i-1/2}) \over 
4(r^2_{i+1/2} + r_{i+1/2}r_{i-1/2} + r^2_{i-1/2})^2}\right]
\end{eqnarray}
where ${g_p}_i^n$ is the gradient of the pressure in cell $i$ at time-step
$n$.

The fluxes $\vec{F}$ are calculated from the primitive variables
extrapolated to the cell edges using non-linear averaging of the
gradients (e.g.\ van Leer 1977; Downes \& Ray 1998), making the code second 
order in space.  A non-linear Riemann solver is used in the presence of 
strong rarefactions, and otherwise a linear Riemann solver is used in order 
to save computational time.  The flux terms $\vec{F}_{i\pm1/2}^{n+1/2}$ are 
calculated using a first order Godunov-scheme from the conditions at 
time $n$.  This makes the code second order in time.  

The details of the linear and nonlinear Riemann solvers follow closely
the discussions of Falle \shortcite{falle91} and Falle \& Komissarov 
\shortcite{fallekomissarov96}.

The code also contains an option for hierarchical grid refinement.  The
algorithm used for this is loosely based on that of Khokhlov 
\shortcite{khokhlov}.  However, this option is not used in the work
presented here, so we will not discuss this further.

\subsection{Tests of the hydrodynamical code}
\label{test}
The code used in this work has been rigorously tested against, in particular, 
results presented in Falle \& Komissarov (1996). The code reproduces their 
published results.  Below we show three examples of these tests for 
completeness.

\subsubsection{Test 1: Shock-tube problem with $\gamma = {4 \over 3}$}

\begin{figure}
\begin{center}
\leavevmode
\epsfxsize = 240pt \epsfbox{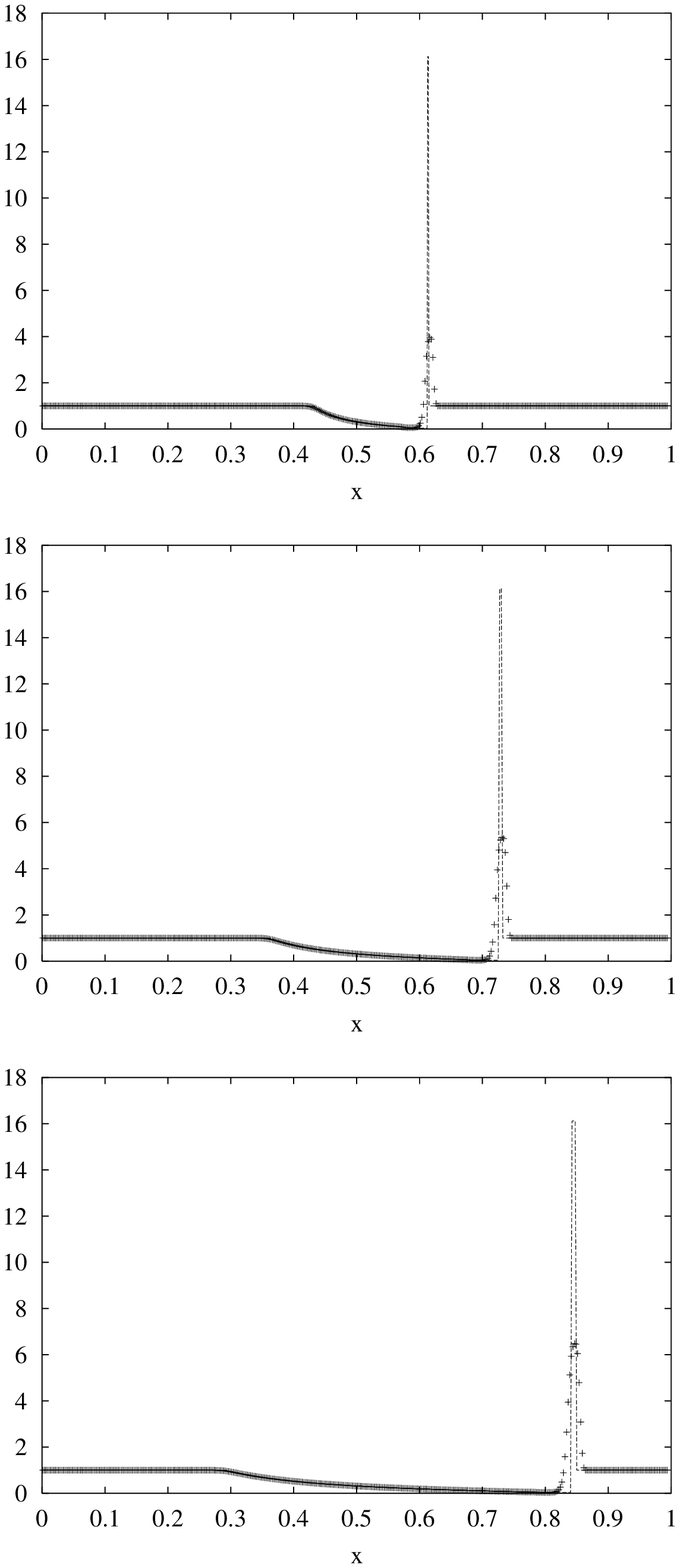}
\caption{\label{fig_test1}  Plots of the proper density at $t=$0.12,
0.24 and 0.36.  The points are the numerical solution, while the line is the 
exact solution.  The rarefaction is tracked extremely well, while the
shock is slightly smeared.  The contact is more smeared, as expected.
See text.} 

\end{center}
\end{figure}

In this test we use the following initial conditions:
\begin{eqnarray*}
\rho & = & 1 \\
u & = & 0 \\
p & = & \left\{\begin{array}{ll}
                    10^3 & \mbox{on $x\leq0.5$} \\
                    10^{-2} & \mbox{on $x > 0.5$}
                    \end{array}
            \right.
\end{eqnarray*}
with gradient zero boundary conditions on $x=0$ and $x=1$.  This
calculation was done for a uniform grid with 400 cells.  The results are
shown in figure \ref{fig_test1}, along with the exact solution.  The
plots are of the proper density, and are at different times.  It can be
seen that, while the contact discontinuity becomes quite smeared (as
expected, since it is a linear wave) the shock and rarefaction are
treated relatively well.

In particular, the position of both the shock and the rarefaction are
very well calculated, indicating that both the linear and non-linear
Riemann solvers are working well.

\subsubsection{Test 2: Shock-tube problem for a Synge gas}

\begin{figure}
\begin{center}
\leavevmode
\epsfxsize = 240pt \epsfbox{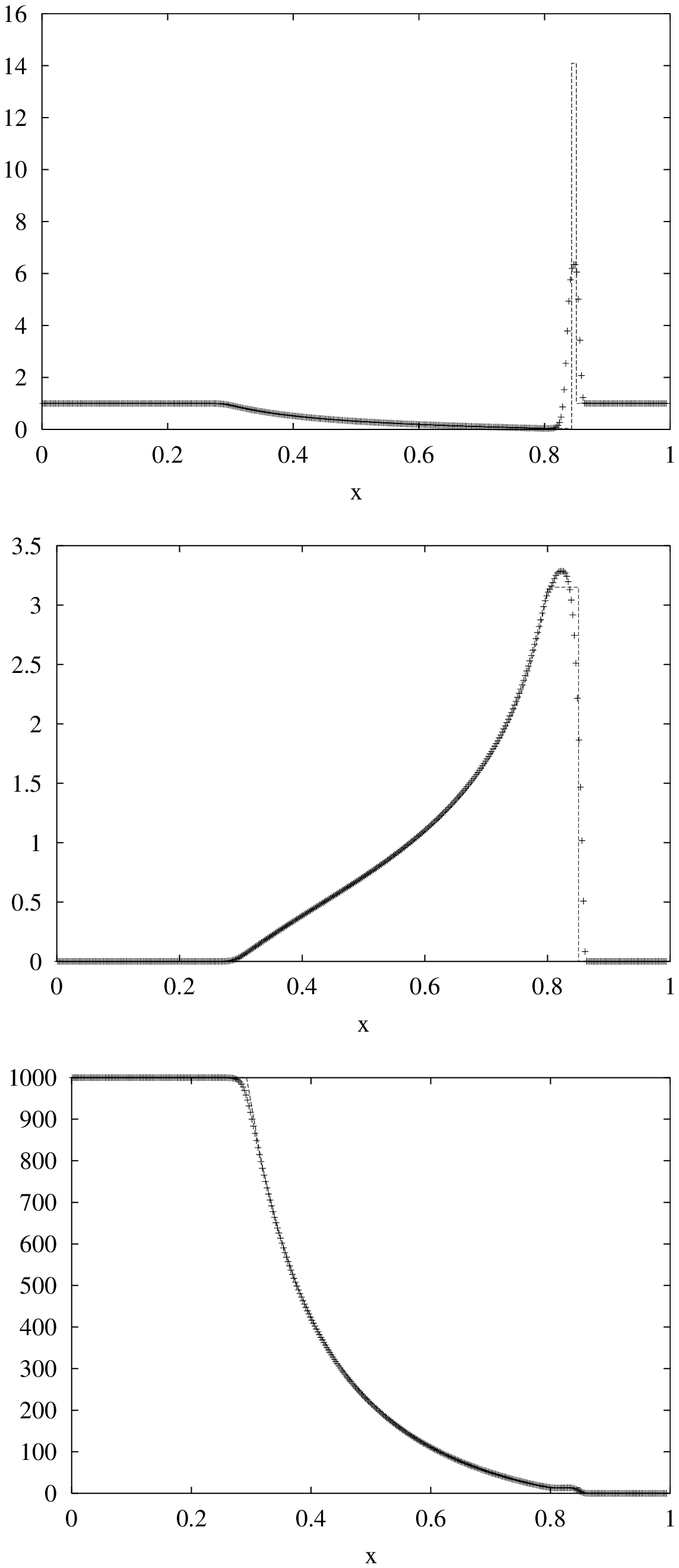}
\caption{\label{fig_test2}  Plots of density, spatial component of the 
4-velocity and pressure (top to bottom) for the shock-tube test.  The points 
are the numerical solution, while the line is the exact solution.  The 
numerical solution is virtually indistinguishable from that in Falle \& 
Komissarov \shortcite{fallekomissarov96}.}

\end{center}
\end{figure}

In this test we use the following initial conditions:
\begin{eqnarray*}
\rho & = & 1 \\
u & = & 0 \\
p & = & \left\{\begin{array}{ll}
                    10^3 & \mbox{on $x\leq0.5$} \\
                    10^{-2} & \mbox{on $x > 0.5$}
                    \end{array}
            \right.
\end{eqnarray*}
with gradient zero boundary conditions on $x=0$ and $x=1$.  This
calculation was done for a uniform grid with 400 cells.  The
results are shown in figure \ref{fig_test2}, along with the exact
solution.  Again, it can be seen that the numerical results match the
analytic results reasonably well, although numerical dissipation does
affect the contact discontinuity quite significantly.

\subsubsection{Test 3: Spherically symmetric expanding wave}
\label{code_test_3}

\begin{figure}
\begin{center}
\leavevmode
\epsfxsize = 240pt \epsfbox{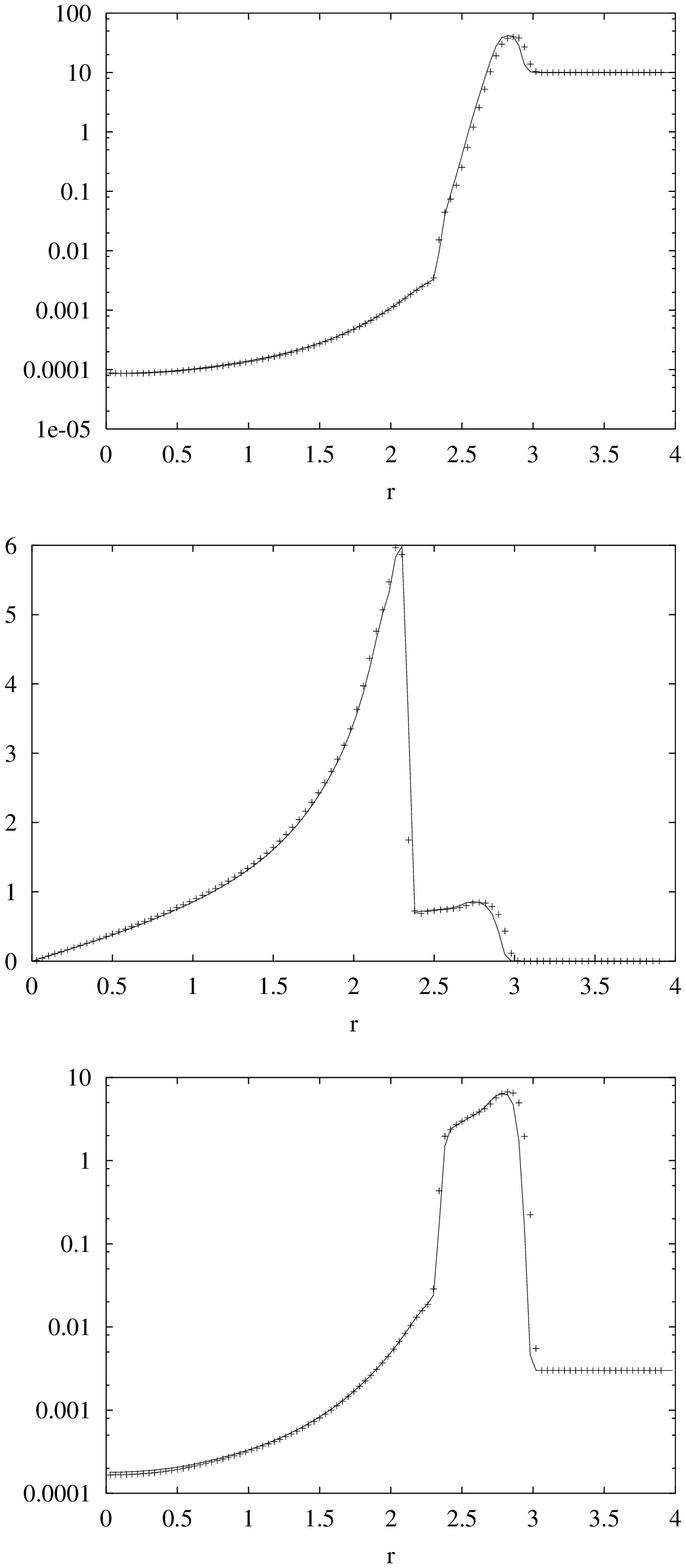}
\caption{\label{fig_test3}  Plots of density, spatial component of the 
4-velocity and pressure (top to bottom).  The points show our solution, while 
the line shows the solution from the code of Falle \& Komissarov 
\shortcite{fallekomissarov96}.  It can be seen that two compare very well.}

\end{center}
\end{figure}

This test is very similar to the simulations used in this work.
Therefore it is an interesting test to run.  The conditions are as
follows:
\begin{eqnarray*}
\rho & = & 10 \\
u & = & 0 \\
p & = & \left\{\begin{array}{ll}
                    10^3 & \mbox{on $r\leq0.5$} \\
                    3\times10^{-3} & \mbox{on $x > 0.5$}
                    \end{array}
            \right.
\end{eqnarray*}
The grid spans $r=0$ to $r=4$ with 100 uniformly spaced cells.  The
results obtained are very similar to those given in Falle \& Komissarov
\shortcite{fallekomissarov96} and are shown in figure \ref{fig_test3}.

Overall, then, the code reproduces published results very well.  One
additional test was performed.  This was a test of the code for the
non-relativistic Sedov blastwave.  The code successfully reproduced the
behaviour of the radius of the blastwave (with $r \propto t^{2 \over
5}$).

It is noted in Sect.\ \ref{hydro_results} that the simulations performed
here give entropy errors in some parts of the fluid.  These errors arise
when the reflected rarefaction (see Sect.\ \ref{hydro_disc}) has almost
caught up with the reverse shock.  In this case there is an extremely
strong rarefaction, with differences of the Lorentz factor of the fluid
of 30 -- 40 between adjacent grid cells, just behind the reverse shock.
These very extreme conditions result in negative pressures being
produced (in the grid cell, not in the Riemann solver), subsequently 
resulting in entropy errors being apparent in the solution.  

As stated in Sect.\ \ref{hydro_results}, however, these errors occur in
a region of the flow where the density of emitting particles is
relatively low, and the errors do not affect the simulated emission
appreciably.

\label{lastpage}

\end{document}